\documentclass{article}
\usepackage{graphicx}
\usepackage{amsmath}

\newtheorem{theorem}{Theorem}
\newtheorem{acknowledgement}[theorem]{Acknowledgement}

\input{tcilatex}

\begin{document}

\title{\textbf{Gauge field divergences in the light-front}}
\author{B.M.Pimentel$^{a}$, J.H.O.Sales$^{a}$ and Tobias Frederico$^{b}$ \\
$^{a}$Instituto de F\'{i}sica Te\'{o}rica-UNESP, 01405-900 S\~{a}o Paulo,
Brazil.\\
$^{b}$Instituto Tecnol\'{o}gico de Aerona\'{u}tica, CTA, 12228-900\\
S\~{a}o Jos\'{e} dos Campos, Brazil.}
\maketitle

\begin{abstract}
The Bethe-Salpeter equation for ground state of two fermions exchanging a
gauge boson presents divergences in the momentum transverse, even in the
ladder aproximation projected in light-front. Gauge theories with
light-front gauge also present the difficulty associated to the
instantaneous term of the propagator of a system composed by fermions
bosons-exchange interaction. We used a prescription that allowed an
apropriate description of the singularity in the propagator of the gauge
boson in the light-front.
\end{abstract}

\section{Light-Front Dynamics: Definition}

Beginning from Dirac's idea \cite{dirac} of representing the dynamics of the
quantum system at ligth-front times $x^{+}=t+z$, we derive the Green's
function from the covariant propagator that evolutes the system from one
light-front hyper-surface to another one. The light-front Green's function
is the probability amplitude for an initial state at $x^{+}=0$ do evolvey to
a final state in the Fock-state at some $x^{+}$, where the evolution
operator is defined by the light-front Hamiltonian \cite{Kogut}

\section{The Scalar Field Propagator}

The Feynman propagator for the scalar field is 
\begin{equation}
S(x^{\mu })=\int \frac{d^{4}k}{\left( 2\pi \right) ^{4}}\frac{ie^{-ik^{\mu
}x_{\mu }}}{k^{2}-m^{2}+i\varepsilon }.  \label{1}
\end{equation}
and in terms of light-front variables \cite{jhs2002}, we have 
\begin{equation}
S(x^{+})=\frac{1}{2}\int \frac{dk^{-}dk^{+}dk^{\perp }}{\left( 2\pi \right) }%
\frac{ie^{\frac{-i}{2}k^{-}x^{+}}}{k^{+}\left( k^{-}-\frac{k_{\perp
}^{2}+m^{2}-i\varepsilon }{k^{+}}\right) }.  \label{2}
\end{equation}

The Fourier transform of the single boson state propagator to the in the
light-front time is giver by: 
\begin{equation}
\widetilde{S}(k^{-})=\int dk^{+}dk^{\perp }\frac{i}{k^{+}\left( k^{-}-\frac{%
k_{\perp }^{2}+m^{2}-i\varepsilon }{k^{+}}\right) }.  \label{3}
\end{equation}

\section{Fermion Field}

Let $S_{\text{F}}$ denote fermion field propagator in covariant theory 
\begin{equation}
S_{\text{F}}(x^{\mu })=\int \frac{d^{4}k}{\left( 2\pi \right) ^{4}}\frac{i(%
\rlap\slash k_{\text{on}}+m)}{k^{2}-m^{2}+i\varepsilon }e^{-ik^{\mu }x_{\mu
}},  \label{4}
\end{equation}
where $\rlap\slash k_{\text{on}}=\frac{1}{2}\gamma ^{+}\frac{(k^{\perp
})^{2}+m^{2}}{k^{+2}}+\frac{1}{2}\gamma ^{-}k^{+}-\gamma ^{\perp }k^{\perp }$%
. Using light-front variables in Eq.(\ref{4}), we have 
\begin{equation}
S_{\text{F}}(x^{+})=\frac{i}{2}\int \frac{dk^{-}dk^{+}dk^{\perp }}{\left(
2\pi \right) }\left[ \frac{\rlap\slash k_{on}+m}{k^{+}\left(
k^{-}-k_{on}^{-}+\frac{i\varepsilon }{k^{+}}\right) }+\frac{\gamma ^{+}}{%
2k^{+}}\right] e^{\frac{-i}{2}k^{-}x^{+}}.  \label{5}
\end{equation}

We note that for the fermion field, light-front propagator differs from the
Feynmam propagator by an instantaneous propagator.

\section{Gauge Boson Propagator}

Let $S^{\mu \nu }$gauge propagator, 
\begin{equation}
S^{\mu \nu }(x^{\mu })=\int \frac{d^{4}k}{\left( 2\pi \right) ^{4}}\frac{%
ie^{-ik^{\mu }x_{\mu }}}{k^{2}+i\varepsilon }\left[ \frac{-nkg^{\mu \nu
}+n^{\mu }k^{\nu }+n^{\nu }k^{\mu }}{nk}\right] ,  \label{6}
\end{equation}
where we choose the light-front gauge $A^{+}=0$, $n^{\mu }=(1,0,0,-1)$ and
the metric tensor is given from \cite{Kogut}$.$

The light-front components (\ref{6}) can be as written $%
S^{+-}=S^{-+}=S^{++}=S^{+\perp }=0$ and 
\begin{equation}
S^{--}=4\frac{ik^{-}}{k^{+}(k^{2}+i\varepsilon )},\text{ }S^{-\perp
}=S^{\perp -}=2\frac{ik^{\perp }}{k^{+}(k^{2}+i\varepsilon )},\text{ }%
S^{\perp \perp }=-1\frac{i}{k^{2}+i\varepsilon }  \label{7a}
\end{equation}

\section{Interaction in First Order}

We consider the fermion-antifermion system in the light-front with one-gauge
boson exchange ($A^{+}=0$), for which the interaction Lagrangian density is
given by 
\begin{equation}
\mathcal{L}_{I}=g\overline{\Psi }_{1}\gamma _{\mu }A^{\mu }\Psi _{1}+g%
\overline{\Psi }_{2}\gamma _{\nu }A^{\nu }\Psi _{2}.  \label{8}
\end{equation}
The fermion corresponds to the field $\Psi $ with rest masses $m$ and the
exchanged gauge boson to the field $A^{\mu }$ with mass $\mu =0.$ The
coupling constant is $g.$

The perturbative correction to the two-body propagator which comes from the
\ exchange of one intermediate virtual boson, is 
\begin{eqnarray}
\Delta S_{g^{2}}(x^{+}) &=&\left( ig\right) ^{2}\int d\overline{x}_{1}^{+}d%
\overline{x}_{2}^{+}S_{k^{\prime }}(x^{+}-\overline{x}_{1}^{+})(\gamma _{\mu
})S_{k}(\overline{x}_{1}^{+})  \label{9} \\
&&S^{\mu \nu }(\overline{x}_{2}^{+}-\overline{x}_{1}^{+})S_{p}(x^{+}-%
\overline{x}_{2})(\gamma _{\nu })S_{p^{\prime }}(\overline{x}_{2}^{+}). 
\notag
\end{eqnarray}
The intermediate boson propagates between the time interval $\overline{x}%
_{2}^{+}-\overline{x}_{1}^{+}.$ The labels in the particle propagators $k$
and $p$ indicates initial and $k^{\prime }$ and $p^{\prime }$ final states.

Performing the Fourier transform from $x^{+}$ to $P^{-}$ and for the total
kinematical momentum $P^{+}$, which we choose positive, and $P^{\perp }$.
The double integration in $k^{-}$ is performed analytically in Eq.(\ref{10a}%
), 
\begin{eqnarray*}
\Delta S_{g^{2}}(P^{-}) &=&\frac{-\left( ig\right) ^{2}i}{(4\pi )^{2}}\int 
\frac{dk^{-}dk^{\prime ^{-}}}{k^{+}k^{\prime ^{+}}(P^{+}-k^{\prime
+})(P^{+}-k^{+})} \\
&&\left\{ \frac{\rlap\slash k_{on}^{\prime }+m}{\left( k^{\prime
-}-k_{on}^{\prime -}+\frac{i\varepsilon }{k^{\prime +}}\right) }\right. 
\begin{array}{c}
\gamma _{-}
\end{array}
\frac{\rlap\slash k_{on}+m}{\left( k^{-}-k_{on}^{-}+\frac{i\varepsilon }{%
k^{+}}\right) } \\
&&\frac{4\left( k^{-}-k^{\prime -}\right) }{(q^{+})^{2}\left(
k^{-}-k^{\prime -}-q_{on}^{-}+\frac{i\varepsilon }{q^{+}}\right) }\frac{%
\rlap\slash p_{on}^{\prime }+m}{\left( p^{\prime -}-p_{on}^{\prime -}+\frac{%
i\varepsilon }{p^{\prime +}}\right) } \\
&& 
\begin{array}{c}
\gamma _{-}
\end{array}
\frac{\rlap\slash p_{on}+m}{\left( p^{-}-p_{on}^{-}+\frac{i\varepsilon }{%
p^{+}}\right) }+
\end{eqnarray*}

\begin{eqnarray}
&&+\frac{\rlap\slash k_{on}^{\prime }+m}{\left( k^{\prime -}-k_{on}^{\prime
-}+\frac{i\varepsilon }{k^{\prime +}}\right) }  \notag \\
&& 
\begin{array}{c}
\gamma _{-}
\end{array}
\frac{\rlap\slash k_{on}+m}{\left( k^{-}-k_{on}^{-}+\frac{i\varepsilon }{%
k^{+}}\right) }\frac{2\left( k^{\perp }-k^{\prime \perp }\right) }{%
(q^{+})^{2}\left( k^{-}-k^{\prime -}-q_{on}^{-}+\frac{i\varepsilon }{q^{+}}%
\right) }  \notag \\
&&\frac{\rlap\slash p_{on}^{\prime }+m}{\left( p^{\prime -}-p_{on}^{\prime
-}+\frac{i\varepsilon }{p^{\prime +}}\right) } 
\begin{array}{c}
\gamma _{\perp }
\end{array}
\frac{\rlap\slash p_{on}+m}{\left( p^{-}-p_{on}^{-}+\frac{i\varepsilon }{%
p^{+}}\right) }+\left[ \gamma _{\perp }\rightarrow \gamma _{-}\right] + 
\notag \\
&&\frac{\rlap\slash k_{on}^{\prime }+m}{\left( k^{\prime -}-k_{on}^{\prime
-}+\frac{i\varepsilon }{p^{\prime +}}\right) } 
\begin{array}{c}
\gamma _{\perp }
\end{array}
\frac{\rlap\slash k_{on}+m}{\left( k^{-}-k_{on}^{-}+\frac{i\varepsilon }{%
p^{+}}\right) }  \notag \\
&&\frac{(-1)}{(q^{+})^{2}\left( k^{-}-k^{\prime -}-q_{on}^{-}+\frac{%
i\varepsilon }{q^{+}}\right) }  \label{10a} \\
&&\frac{\rlap\slash p_{on}^{\prime }+m}{\left( p^{\prime -}-p_{on}^{\prime
-}+\frac{i\varepsilon }{p^{\prime +}}\right) } 
\begin{array}{c}
\gamma _{\perp }
\end{array}
\left. \frac{\rlap\slash p_{on}+m}{\left( p^{-}-p_{on}^{-}+\frac{%
i\varepsilon }{p^{+}}\right) }\right\} ,  \notag
\end{eqnarray}

\section{Conclusion}

From equation (\ref{10a}), we verified the existence of singularity in the
components $(--)$ and $(-\perp )$ in the coordinate $q^{+}=k^{+}-k^{\prime
+}.$ We hoped to remove those singularity using the technique of
displacement $\left( \delta ^{+}\right) $ of the pole of the phase space in $%
q^{+}$ \cite{jhs97}.

\begin{acknowledgement}
T.F and B.M. Pimentel thank to CNPq for partial support. J.H.O. Sales is
supported by FAPESP/Brazil pos-doctoral fellowship. We acknowledge
discussions with J.F.Libonati and J.Messias.
\end{acknowledgement}

\end{document}